# Making the cut: end effects and the benefits of slicing

Bharath Antarvedi Goda[a], David Labonte[b], Mattia Bacca[a*]

[a]*Mechanical Engineering Department, Institute of Applied Mathematics, School of Biomedical Engineering, University of British Columbia, Vancouver BC V6T1Z4, Canada*
[b]*Bioengineering Department, Imperial College London, FRXG+G3, England (UK)*

[*]Corresponding author. *E-mail address:* mbacca@mech.ubc.ca

**Abstract**

Cutting mechanics in soft solids have been a subject of study for several decades, an interest fuelled by the multitude of its applications, including material testing, manufacturing and biomedical technology. Wire cutting is the simplest model system to analyse the cutting resistance of a soft material. However, even for this simple system, the complex failure mechanisms that underpin cutting are still not completely understood. Several models that connect the critical cutting force to the radius of the wire and the key mechanical properties of the cut material have been proposed. An almost ubiquitous simplifying assumption is a state of plane (and anti-plane) strain in the material. In this paper we show that this assumption can lead to erroneous conclusions, because even such a simple cutting problem is essentially three-dimensional. A planar approximation restricts the analysis to the stress distribution in the mid plane. However, through finite element modelling, we reveal that the maximal tensile stress – and thus the likely location of cut initiation – is in fact located in the front plane. Friction reduces the magnitude of this stress, but this detrimental effect can be counteracted by large "*slice-to-push*" (shear-to-indentation) ratios. The introduction of these "*end effects*" helps reconcile a recent controversy around the role of friction in wire cutting, for it implies that slicing can indeed reduce required cutting forces, but only if the slice-push ratio and the friction coefficient are sufficiently large. Material strain-stiffening reduces the critical indentation depth required to initiate the cut further, and thus needs to be considered when cutting non-linearly elastic materials.

## 1. Introduction

The mechanical problem of soft solid cutting gained significant attention in the last few decades, thanks to its numerous applications in material testing (Lake1978), (Gent1996), (Atkins2009), (Patel2009), manufacturing (McGee2013), (Duenser2020), (Kamyab1998), (Goh2005), and medical technology (Sugano2013), (Liu2021-a). Despite the widespread importance of this class of materials, the complex mechanisms involved in cutting them are still poorly understood. Cutting involves the initiation and propagation of a crack via contact loading. Stiff materials typically fail at infinitesimally small deformations. Soft solids, however, undergo large non-linear elastic deformations prior to failure. (Goh2005), (Fregonese 2021), (Fregonese 2023) suggested to distinguish two stages in cutting (puncturing) soft solids: (i) an indentation phase, prior to crack nucleation in the material, and (ii) steady state cutting, characterized by a continuous propagation of the crack at an approximately constant force. Most previous investigations addressed the mechanics of steady state cutting using energy arguments (Kamyab1998), (Goh2005), (Fregonese 2022). Less attention has been paid to the complex mechanism involved in cut initiation, which depend on the stress distribution in the material, and material-specific failure mechanisms. What determines the force required to initiate a cut?

Empirically, cutting can be easier when contact forces are transmitted via a combination of normal and out-of-plane displacements, *i.e.*, when solids are "sliced" (Atkins2004, Reyssat2012, Liu2021). To explain this observation, (Reyssat2012) studied the cutting of agar gels with a cylindrical wire. For shallow indentation depths, the contact forces transmitted by the wire predominantly generate compressive stresses. Tensile stresses, which can promote crack nucleation, emerge close to contact interface only at large indentation depths (Reyssat2012). However, the stress distribution under the wire changes when normal displacements are combined with shear displacements. During such "shear cutting", tensile stresses can emerge at shallower indentation depths, and cut initiation thus requires less effort (Reyssat2012). The positive effect of shear cutting consequently relies on the transmission of contact shear stresses between the wire and the specimen via friction. However, friction also introduces a barrier to crack nucleation, because it constrains the lateral displacement of material underneath the wire, required for tensile stresses (Liu2021). Friction thus appears to play a dual role, and can either hinder or promote crack nucleation. Based on a plane strain and anti-plane shear analysis, (Liu2021) concluded that, for a given indentation depth, maximum tensile stresses always occur in the frictionless condition, where the material is "free" to slide. One may thus surmise that the easiest cut occurs in the absence of friction, and that wire lubrication may be preferable to slicing via shear cutting. Under which conditions is it easiest to nucleate a crack?

Previous works modeled indentation as an in-plane effect and sliding as an anti-plane effect, and considered the stress distribution in a mid-plane of the sample, making this analysis effectively two-dimensional. Here, we revisit this assumption, and consider the three-dimensional stress-distribution which results from shear cutting also in the front- and back-plane of the cut specimen. The location of maximum stresses in the sample is identified and the role of frictional shearing and strain stiffening are carefully studied.

## 2. Mechanical analysis of wire cutting

To investigate the mechanics of wire-cutting, we perform a three-dimensional stress analysis of a parallelepiped specimen indented by a rigid wire (Figure 1a). The wire has a circular cross-section of radius $R$, and indents the specimen's top surface to a depth $d_y$, along the *y*-direction, while simultaneously displacing along the *Z*-axis (wire axis) by a distance $d_z$. We describe the ratio of horizontal to vertical wire displacement with the *shearing* angle $\theta$

$$tan\theta = \frac{d_z}{d_y} \qquad (1)$$

Thus, $\theta = 0$ implies pure indentation, with $d_z = 0$, while $\theta \to 90°$ implies $d_z \gg d_y$, *i.e.*, maximum shear (Figure 1a). We define the displacement ratio $d_z/d_y$ as the *slice-to-push ratio*. In our simulations, we fixed $d_y$ and vary $d_z = d_y tan\theta$. We assume the deformation of the cut material is quasi-static, so that inertia and rate-dependent material behavior can be neglected. To appropriately capture the non-linear strain-stiffening behavior that characterizes many elastomers and soft biological materials, we adopt a 1-term Ogden strain energy density functional (Ogden, 1972)

$$\psi = \frac{2\mu}{\alpha^2}(\lambda_1^\alpha + \lambda_2^\alpha + \lambda_3^\alpha - 3) \quad (2)$$

where $\mu$ and $\alpha$ are the shear modulus and the dimensionless strain-stiffening coefficient of the material, respectively ($\alpha = 2$ yields Neo-Hookean behavior), and $\lambda_i$ are the principal stretches.

The Cauchy stress along the principal direction 1, is

$$\sigma_1 = \lambda_1 \frac{\partial \psi}{\partial \lambda_1} - p \quad (3)$$

where $p$ is the hydrostatic pressure, *i.e.*, a Lagrange multiplier enforcing incompressibility.

To determine the stress field, we perform a finite element analysis (FEA) with the commercial software Abaqus. To minimise specimen size effects, we choose its dimensions $L_x$, $L_y$, and $L_z$ such that they are much larger than $R$, $d_y$ and $d_z$ (Figure 1a). Three regions of interest may be defined: (i) the *front-face*, (ii) the *mid-plane*, and (iii) the *back-face* (Figure 1a), where (ii) and (iii) differ only for $\theta > 0$. The displacements at the vertical boundaries of the specimen are fixed in direction (Figure 1b), which effectively reduces the dimension $L_x$ as required to avoid size-effects. The displacements at the bottom were fixed in the direction *Y* to simulate a rigid supporting plane underneath the specimen (Figure 1b). By virtue of the symmetry with respect to the *Y-Z* plane, we study a half specimen and adopt symmetric boundary conditions in all calculations (Figure 1b). The mesh is refined near the contact region, to capture stress gradients, and coarsens toward the external boundaries of the specimen (Figure 1b). We use C3D8H hexahedral hybrid elements to describe the incompressible behavior of the cut material. The wire is modeled using R3D4 rigid elements. The contact interactions between the wire and the specimen are described via

$$\tau_c \leq \tau_f \quad (4a)$$

with $\tau_c$ the contact shear stress, and $\tau_f$ the maximum shear stress dictated by static friction. The maximum sustainable shear stress follows from Coulomb's law as

$$\tau_f = \zeta P_c \quad (4b)$$

where $\zeta$ is the static friction coefficient, and $P_c$ is the contact pressure. For simplicity, we assume that both static and dynamic friction are adequately described by Eq. (4).

We analyze the distribution of the dimensionless Cauchy stress $\sigma/\mu$ across the (i) front, (ii) mid, and (iii) back planes, and investigate how this distribution is controlled by the indentation depth $d_y$, the shearing angle $\theta$, the friction coefficient $\zeta$, and the strain-stiffening coefficient $\alpha$.

## 3. Results and Discussion

Cutting involves the nucleation and propagation of a crack. There are no universally accepted criteria for fracture nucleation in soft materials; however, mode I fracture propagation and cut initiation is typically prompted by large tensile stresses (Reyssat 2012), (Liu 2021). Accordingly, we analysed the distribution of the maximum principal stress $\sigma_I$ in the specimen. Because the cut is expected to initiate in a symmetry plane Y-Z, orthogonal to the X-axis, where $\sigma_I = \sigma_x$, we focus on the magnitude of $\sigma_x$. As shown in Figure 1a, X, Y, and Z are the coordinates of our reference system, i.e., that of the undeformed specimen.

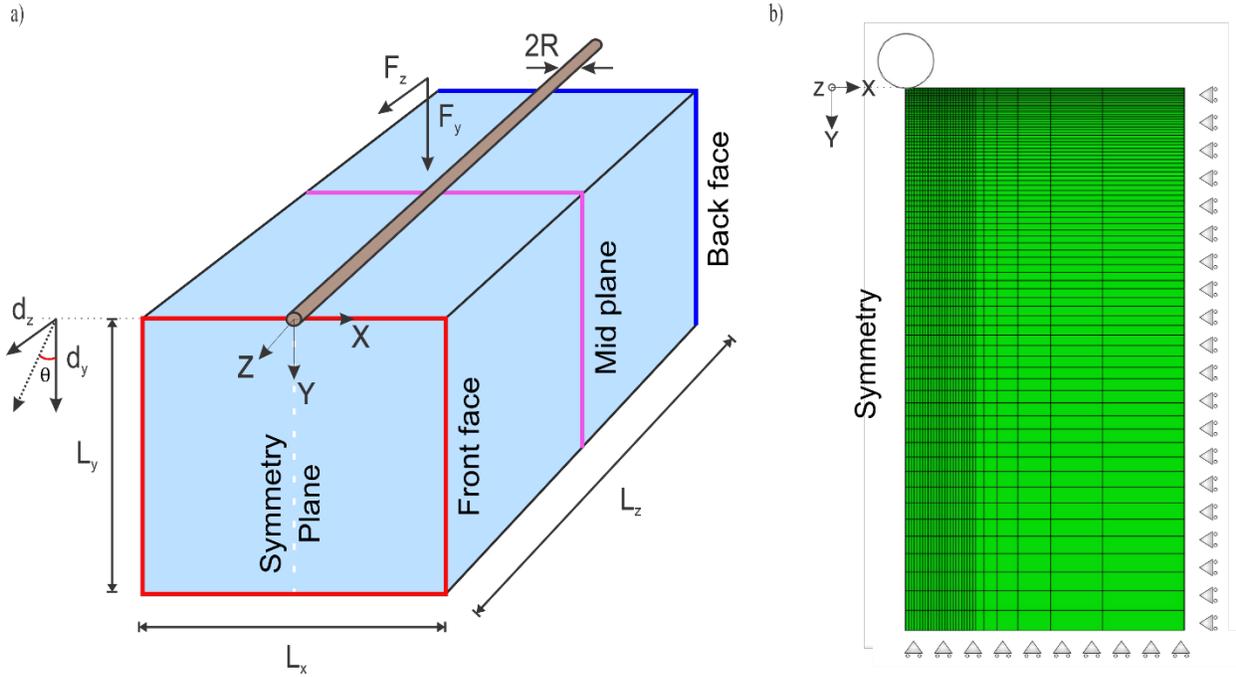

**Figure 1**: a) 3D schematic of the cut specimen and the wire. The analysis focuses on three planes of interest: (I) the *front-face*, (ii) the *mid-plane*, and (iii) the *back-face*. b) Adopted FEA mesh and boundary conditions for the reduced half-model, enabled by the symmetry with respect to the Y-Z plane.

Figure 2 shows the distribution of the dimensionless stress $\sigma_x/\mu$ in the three-dimensional specimen. Tensile stresses are defined as positive. For this calculation, the vertical (indentation) displacement is $d_y/R = 3$, with a shearing angle of $\theta = 0$ (no sliding), a friction coefficient $\zeta = 0.1$, and a strain-stiffening coefficient $\alpha = 2$ (incompressible Neo-Hookean in Eq. (2)). Figure 2a provides the contour plot of the dimensionless stress distribution $\sigma_x/\mu$ in the Y-Z symmetry plane. Figures 2b and 2c plot the distribution of $\sigma_x/\mu$ along the X-axis on the surface of the specimen and along the Y-axis in the symmetry plane, respectively, for both the mid plane (black line) and the front face (blue line). In Figure 2c we highlight the *maximum stress*

$$\sigma_m = \max_Y(\sigma_x) \qquad (5)$$

beneath the contact (solid red circles), and the *surface stress* near the contact

$$\sigma_s = (\sigma_x)_{Y=0} \qquad (6)$$

(solid red triangles, see also Figure 2b). $\sigma_m$ is always positive, i.e., tensile, but $\sigma_s$ is here always negative, i.e., compressive. Because $\sigma_m > \sigma_s$ in all cases, the crack will nucleate consistently in a location at some distance away from the contact. After the crack has nucleated, it must propagate toward the contact to allow the wire to enter the specimen and complete the cut. However, the compressive (negative) stress $\sigma_s$ close to the contact can constitute a barrier to crack propagation. Thus, maximization of both $\sigma_m$ and $\sigma_s$ is instrumental to facilitate cutting. Notably, the maximum stress $\sigma_m/\mu$ at the front face is much larger than that in the mid plane ($\sigma_m/\mu \approx 0.5$ versus $0.1$, Figure 2c), and the surface stress $\sigma_s/\mu$ in the midplane is much more compressive than that in the front face ($\sigma_s/\mu \approx -2$ versus $\approx -0.05$). The cut is consequently more likely to initiate in the front face, and we call this phenomenon the 'end-effect'. The end-effect cannot be identified with a planar analysis. Thus, planar analyses may yield misleading conclusions on the critical force and displacement at which a crack initiates and propagates.

Notably, the maximum tension in the mid plane does not occur at the axis of symmetry ($X = 0$), where cut initiation is usually expected, but at some location on the surface (at $X/R \approx \pm 2.5$ and $Y/R = 0$, Figure 2b). Only with increasing indentation depth, or smaller friction coefficient, does the maximum tension in the mid plane develop in the symmetry plane ($X = 0$). In contrast, the maximum tension always develops in the symmetry plane in the front face, irrespective of indentation depth and friction coefficient. The end effect is thus independent of both.

Consider now the case of wire sliding combined with indentation ($\theta > 0$). In the absence of friction, i.e., $\zeta = 0$, the wire cannot transmit any shear stress to the specimen $\tau_c = 0$. The stress distribution is then independent of $\theta$, and identical to the distribution found for pure indentation $\theta = 0$. As discussed by (Reyssat 2012), interfacial shear stress produces tensile principal stress, which compensates for the compressive stresses created by indentation, but this requires friction ($\zeta > 0$) (Liu 2021).

Next, we varied the shearing angle $\theta$, from 0° (pure indentation) to 90° (maximum shearing), and inspected the corresponding dimensionless maximum stress $\sigma_m/\mu$ (circles) and surface stress $\sigma_s/\mu$ (triangles), for the front (blue) and back faces (red), and for the mid plane (black, Figure 3). The indentation depth was fixed to $d_y/R = 1$ (Figure 3a) or 3 (Figure 3b), and the interfacial and material properties are the same as in Figure 2. Maximum and surface stresses initially increase with the shearing angle $\theta$, but then stabilize above a saturation value, which represents the limit at which the contact shear stress attains the maximum interfacial shear stress defined by Eq. (4a). The critical shear resistance increases with $d_y$, which controls the contact pressure $P_c$ (Eq. (4b)). Consistent with our earlier observations, the maximum tension for $d_y/R = 1$, and for $d_y/R = 3$ occur in the front face, but only when the shearing angle $\theta$ is sufficiently large. Notably, the lowest values for the maximum stress $\sigma_m$ and surface stress $\sigma_s$ always occur in the mid plane. For $d_y/R = 3$ and small $\theta$, the highest tension is situated in the back plane, a surprising result that arises from a competition between the deformation induced by shearing and that due to conservation of volume. The maximum stress $\sigma_m$ increases with $d_y$, as observed in Figure 3, thanks to shearing and conservation of volume, which prompts deformation in direction $Z$ on back and front faces. However, $\sigma_s$ decreases with $d_y$ for almost all the explored cases in Figure 3. This effect is due to friction, which constrains the tensile stretch in the X-Y surface plane required to generate a tensile $\sigma_x$. Because the material needs to expand in the X-direction to preserve the volume under indentation, the frictional forces in direction X generate a compressive surface stress $\sigma_s$. However, this effect disappears for sufficiently large $\theta$ (roughly, for $\theta > 45°$, or $d_z > d_y$ in Figure 3), because of the competition between frictional forces in direction X, prompting a compressive $\sigma_s$, and those in direction Z, created by $d_z$ and prompting tension. The force required for cut initiation can thus be reduced via friction-mediated shear cutting, as discussed by (Reyssat 2012), but this requires a sufficiently large shearing angle $\theta$. This critical shearing angle decreases with increasing indentation depth (Figure 3), and with increasing friction coefficient (Figure 4).

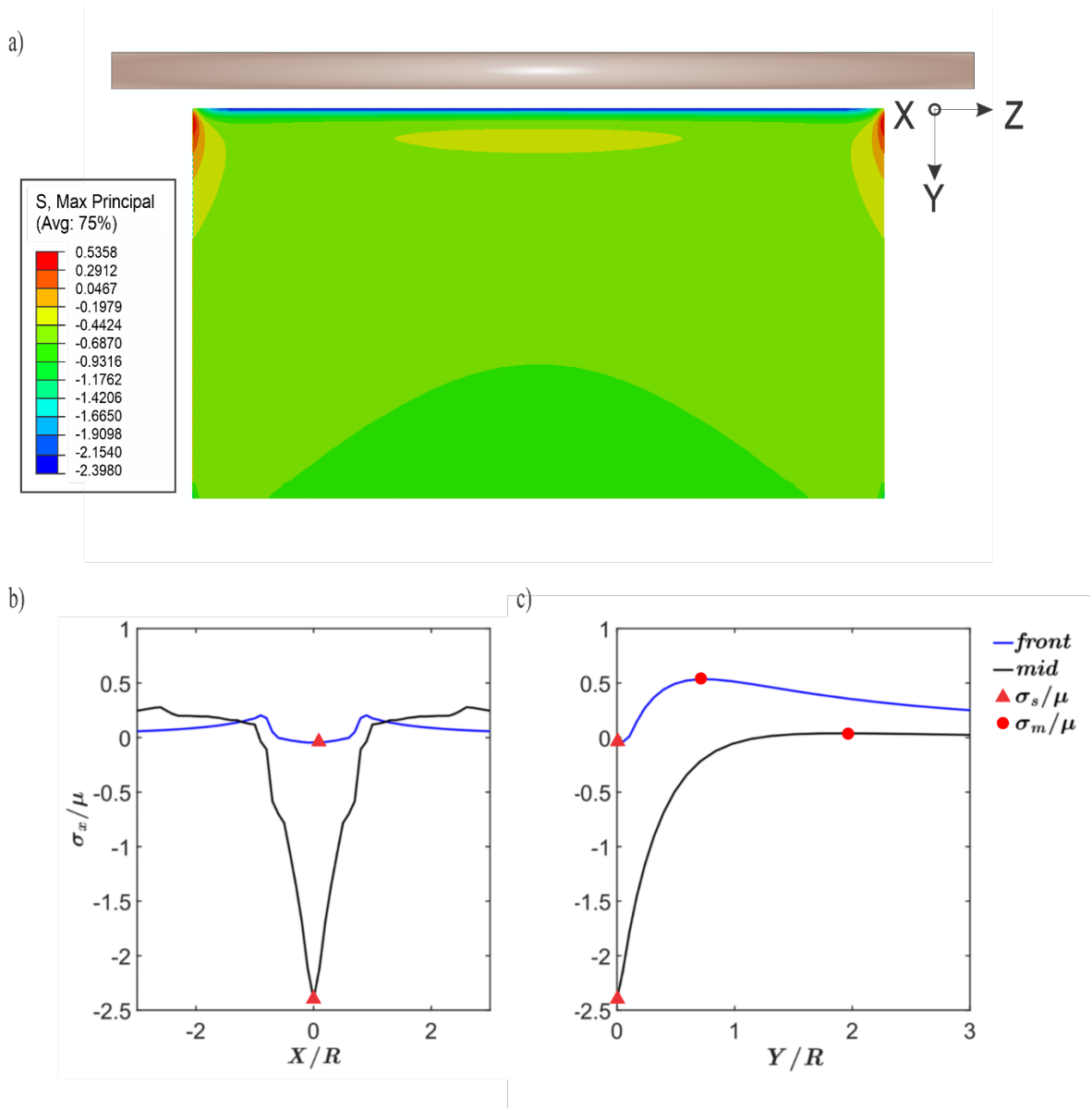

**Figure 2**: Dimensionless Cauchy (true) stress $\sigma_x/\mu$ for the model system shown in Figure 1. *a*) Contour plot of $\sigma_x/\mu$ in the Y-Z symmetry plane. *b*) Distribution of $\sigma_x/\mu$ along the X axis on the surface of the specimen, and *c*) along the Y axis in the symmetry plane, for both the front face (blue lines) and the mid plane (black lines). In the Y-Z symmetry plane (*a* and *c*), $\sigma_x$ corresponds to the maximum principal stress, and the likely

location of crack initiation. We identify a maximum stress $\sigma_m/\mu$ (solid red circles) along Y, and a stress at the contact surface $\sigma_s/\mu$ (solid red triangles). Note that the maximum tensile stress $\sigma_m$ occurs at some distance away from the contact, and is about 5-fold higher at the front face than the mid plane (where maximum $\sigma_x$ is away from the symmetry plane). The stresses directly underneath the contact region, $\sigma_s$, are compressive in both planes, but compression is generally higher in the mid-plane. In this analysis, the indentation depth is $d_y/R = 3$, the shearing angle is $\theta = 0$, the friction coefficient is $\zeta = 0.1$, and the strain-stiffening coefficient is $\alpha = 2$ (neo-Hookean).

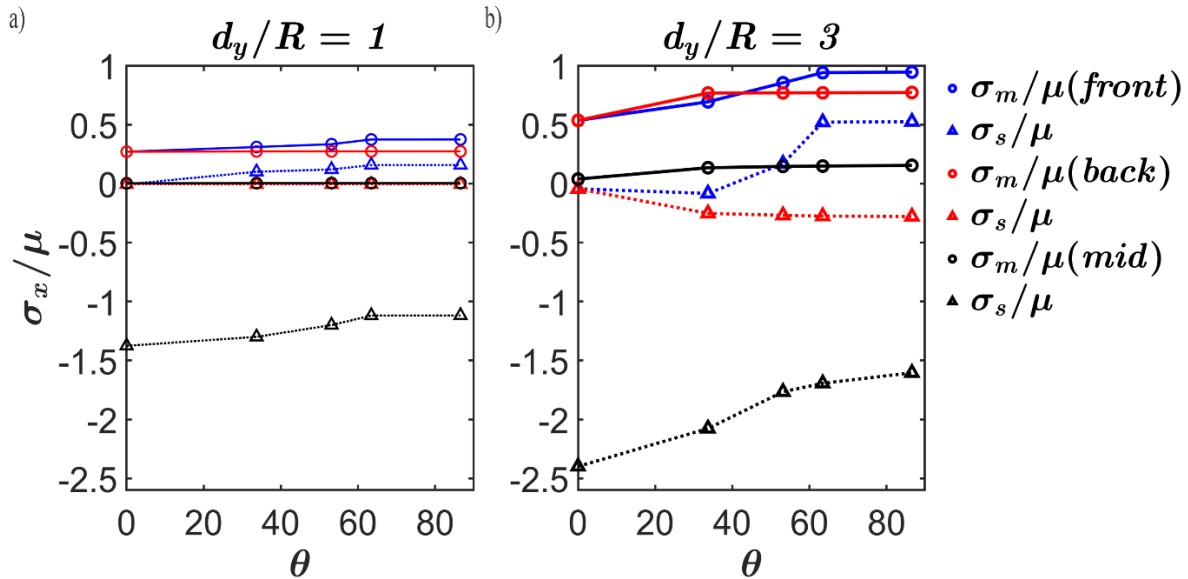

**Figure 3**: Distribution of the maximum (dimensionless) tensile stress $\sigma_m/\mu$ (circles and solid lines) and surface stress $\sigma_s/\mu$ (triangles and dashed lines), in the Y-Z symmetry plane (Figure 2c), versus shearing angle $\theta$ ($\tan\theta = d_z/d_y$) for *front* (blue) and *back* (red) faces, and for *mid* (black) plane. The indentation depth is a) $d_y/R = 1$ and b) $d_y/R = 3$. The friction coefficient is $\zeta = 0.1$, and the strain-stiffening coefficient is $\alpha = 2$ (neo-Hookean) for both *a*) and *b*). Front and back faces always host the highest maximum tension $\sigma_m$, compared to the mid plane, and the front face hosts the maximum tension overall, but only for sufficiently large shearing angle $\theta$. Higher indentation depth $d_y$ induces a higher maximum stress $\sigma_m$ in all cases, and more compressive (smaller) surface stress $\sigma_s$. However, this trend reverses for sufficiently high shearing angles $\theta$, because the tensile stresses induced by frictional forces in the Z-direction dominate over the compressive effect produced by frictional forces in the *X-Y* plane.

In order to explore the effect of friction in mediating compressive vs tensile stresses further (Figure 4a-b), we varied the friction coefficient $\zeta$, from 0 (frictionless), to 0.05 and 0.1, and again consider the dimensionless maximum stress $\sigma_m/\mu$ (circles) and surface stress $\sigma_s/\mu$ (triangles) (Figure 2c) on the front face, as a function of the shearing angle $\theta$. Indentation creates both compressive (blue) and tensile stresses (red) at the front face (Figure 4a-b). However, surface stresses are predominantly compressive under pure indentation, as deformations are localized to the *X-Y* plane. When $\theta \to 0°$, friction penalizes tensile deformations in X at the contact along the symmetry plane (Figure 4a). However, when $\theta \to 90°$, friction relaxes compressive stresses in-plane and promote tensile stresses at the front face via end-effects (Figure 4b). The numerical results are reported in Figure 4c, where $\theta$ goes from 0° to 90°, the indentation depth is $d_y/R = 3$, and the strain-stiffening coefficient is $\alpha = 2$. Surprisingly, for small shearing angles $\theta$ (roughly below 45°, *i.e.*, for $d_z < d_y$), both $\sigma_m$ and $\sigma_s$ decrease with friction ($\zeta$), but for large $\theta$ (45°, or $d_z > d_y$), $\sigma_m$ and $\sigma_s$ increase with $\zeta$. This phenomenon is again explained by the dual role of friction: frictional forces in *X*-direction constrain the lateral expansion of the material required to preserve the volume (Figure 4a), but frictional forces in the *Z*-direction, due to $d_z$, release these constraint and prompt tension (Figure 4b). Both the frictional forces in the *X* and *Z* directions are proportional to the contact pressure $P_c$, and hence to $d_Y$, and to the friction coefficient $\zeta$ (Eq. (4)). We can then conclude that, for small $d_z/d_y$ (small $\theta$), the compressive effect of friction prevails over the tensile one, and maximum tension occurs in frictionless conditions, as proposed by (Liu 2021). Conversely, for large $d_z/d_y$ (large $\theta$) the tensile effect of friction prevails over the compressive one and the highest friction coefficient provides the highest tension, giving merit to shear cutting (Reyssat 2012). This transition can only be observed if the three-dimensionality of the problem is taken into account, as it arises from the *end effect --- the maximum tensile stresses occur at the front face, and not in the mid-plane*. Notably, at large shearing angles $\theta$, the maximum stress $\sigma_m$ appears to be relatively insensitive to the friction coefficient $\zeta$, for $\zeta \geq 0.05$, but the surface stress $\sigma_s$ increases significantly with $\zeta$. This difference arises because the location of the maximum tension $\sigma_m$, lies at some distance from the contact region, whereas the surface stress $\sigma_s$ is located right at the contact, and is thus directly impacted by frictional shear stresses. Notably, very small shearing angles $\theta$ have a negligible effect on the stresses, which highlights that a minimum angle $\theta$ needs to be exceeded to benefit from shear cutting.

The elastic response of the material has thus far been modeled with a neo-Hookean constitutive law, *i.e.*, with a strain-stiffening coefficient of $\alpha = 2$. Let us now consider the effect of strain-stiffening by varying $\alpha$ from 2 (blue lines) to 5 (red lines), as reported in Figure 5. Here, the indentation depth is $d_y/R = 3$ and the friction coefficient is $\zeta = 0.1$. We again study the variation of the dimensionless maximum stress $\sigma_m/\mu$ and surface stress $\sigma_s/\mu$ (Figure 2c) with the shearing angle $\theta$ for the front face. Confirming our previous observations, the benefits of shear cutting can only be appreciated at large $\theta$. Notably, strain stiffening does not appear to affect the surface stress $\sigma_s$ significantly, but a higher $\alpha$ increases $\sigma_m$. This suggests that cutting materials with higher strain-stiffening is easier with shear-cutting: cuts initiate at shallower depths (given that all stresses are proportional to indentation depth). Strain-stiffening increases the elastic stress in the material, and thus amplifies the development of contact pressure $P_c$, which in turn increases the magnitude of frictional forces. The negligible change in $\sigma_s$ suggests that the frictional forces in both *X* and *Z* directions are equally amplified by $\alpha$, while the pronounced change in $\sigma_m$ suggests that maximum tension in the front face is further increased by $\alpha$. This is due to the higher stress produced by strain stiffening in response to the out of plane deformation of the front face.

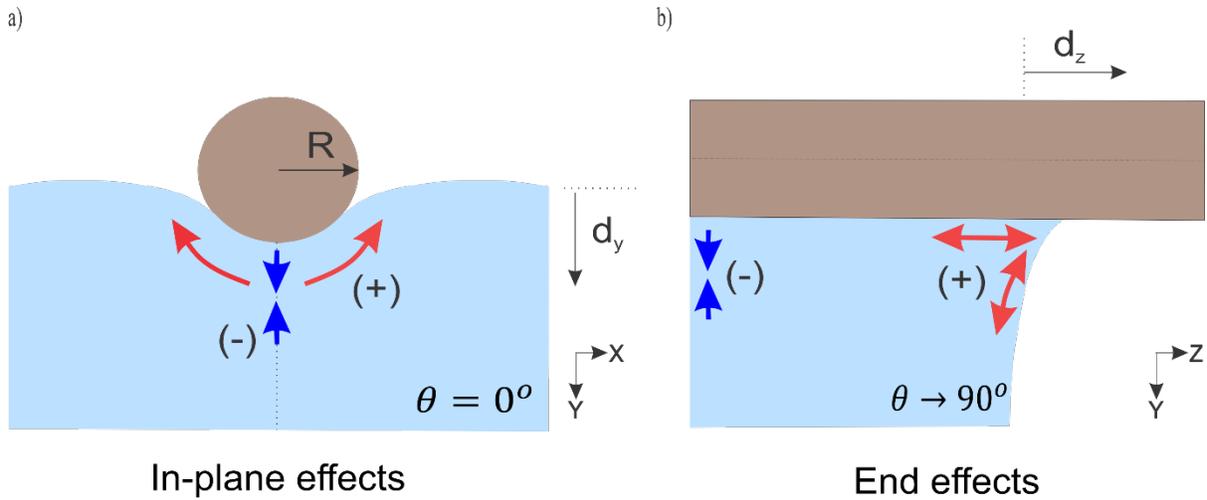

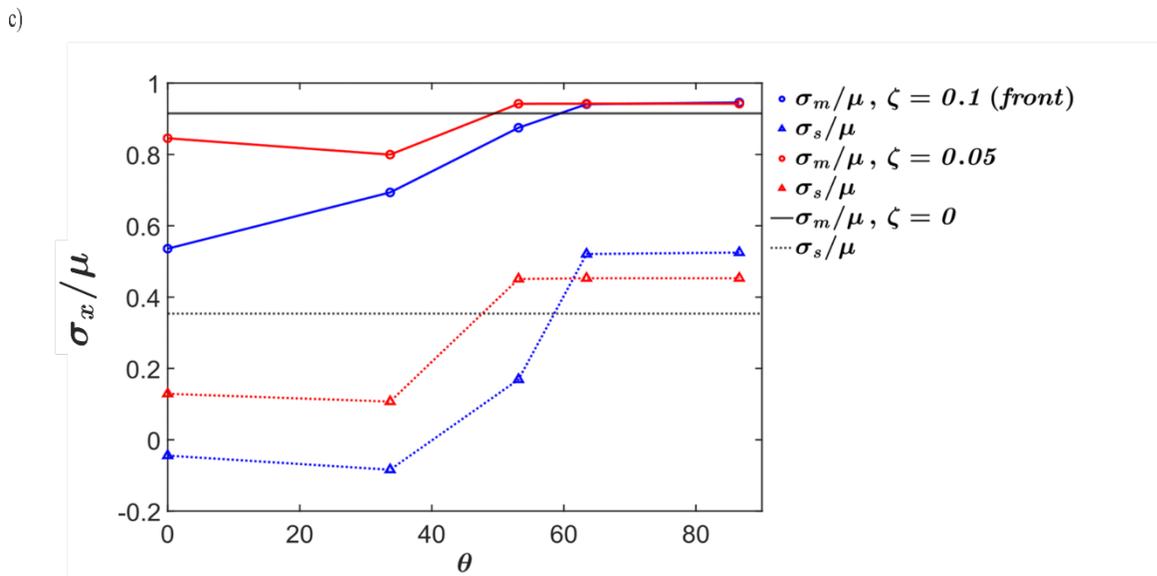

**Figure 4**: *a*) Schematic representation of the dual role of friction: In-plane effects *b*) End effects *c*) Distribution of the maximum (dimensionless) tensile stress $\sigma_m/\mu$ (circles and solid lines) and surface stress $\sigma_s/\mu$ (triangles and dashed lines), in the *Y-Z* symmetry plane (Figure 2c), versus shearing angle $\theta$ ($\tan\theta = d_z/d_y$), for the front face, with a friction coefficient $\zeta$ of $0$ (black), $0.05$ (red), and $0.1$ (blue). The indentation depth is $d_y/R = 3$ and the strain-stiffening coefficient is $\alpha = 2$ (neo-Hookean). In the frictionless case ($\zeta = 0$), the stresses are unaffected by the shearing angle $\theta$, because the wire cannot transmit any shear stress. The addition of friction reduces both $\sigma_m$ and $\sigma_s$ for low shearing angles, because material expansion in the *X*-direction -to conserve volume- prompts a compressive $\sigma_x$ at the contact. In this regime, both stresses are approximately invariant to $\theta$ up until a threshold of $\theta \approx 40°$. For large shearing angles $\theta$, in turn, friction instead increases $\sigma_m$ and $\sigma_s$, with a higher relative increment for $\sigma_s$. This is due to a predominance of frictional forces in the *Z*-direction prompting tension at the contact.

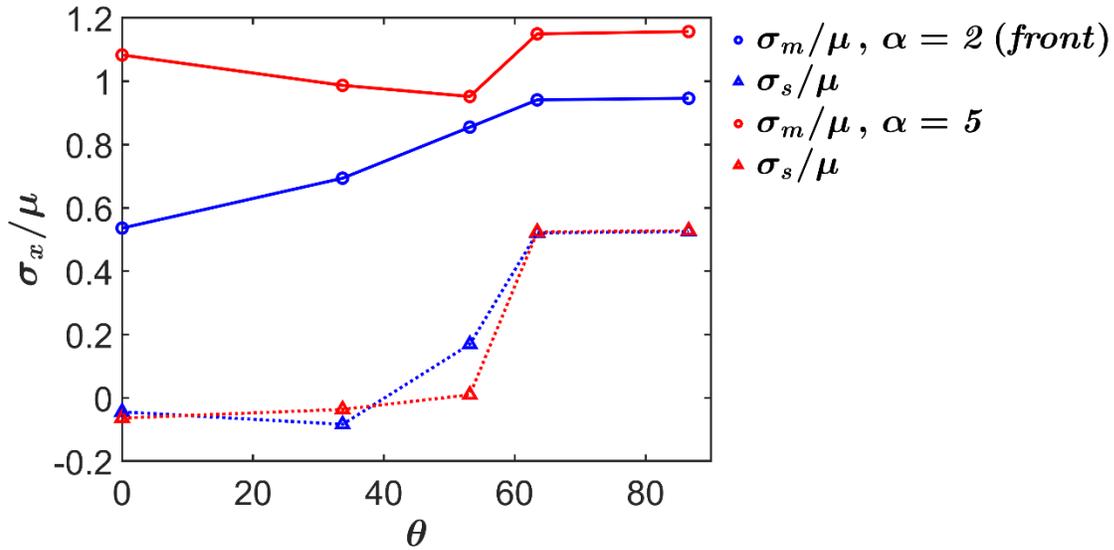

**Figure 5**: Distribution of the maximum dimensionless tensile stress $\sigma_m/\mu$ (circles and solid lines) and surface stress $\sigma_s/\mu$ (triangles and dashed lines), in the *Y-Z* symmetry plane, versus shearing angle $\theta$ ($\tan\theta = d_z/d_y$), for the front face, with strain-stiffening coefficients $\alpha = 2$ (neo-Hookean, blue) and $\alpha = 5$ (red). The friction coefficient is $\zeta = 0.1$, and the indentation depth is $d_y/R = 3$. As observed in the previous figures, shear cutting is beneficial only at sufficiently large $\theta$. Strain stiffening does not affect $\sigma_s/\mu$ significantly, because the increasing contact pressure affects compression in the *X*-direction, and tension in the *Z*-direction similarly. Strain-stiffening does however increase $\sigma_m/\mu$, due to the increased stress produced in response to the out-of-plane deformation of the front face.

## 4. Conclusions

Our study highlights the importance of accounting for the three-dimensionality of the model system when performing a stress analysis in wire cutting. The stress state varies across the thickness of the sample, and the three-dimensional stress distribution can be altered by frictional shearing. A simple analysis based on planar approximations is more convenient, but can result in misleading conclusions, because it cannot identify the '*end effect*', i.e., the occurrence of larger tensile stresses at the front face. Identification of this end effect helps to resolve the controversy around the benefits of friction-mediated `push-and-slice' (shear) cutting. (Reyssat2012) showed that the friction-mediated contact shear stress, created via wire motion along its axis, can lower the normal cutting force, but (Liu2019) argued that frictionless pure indentation results in the lowest cutting force. Our results demonstrate that these two seemingly contradictory conclusions are in fact both correct, each in a limited regime of shearing angles: for small shearing angles (roughly $\theta < 45°$ and $d_z < d_y$), pure indentation represents optimal conditions (Liu2019), but for large shearing angles (roughly $\theta > 45°$ and $d_z > d_y$) wire sliding reduces cutting force requirements (Reyssat2012). In general, the benefits of sliding increase with friction coefficient, contact pressure and strain-stiffening.

*Acknowledgments*

The work was supported by the Human Frontiers Science Program (RGY0073/2020), the Department of National Defense (DND) of Canada (CFPMN1–026), and the Natural Sciences and Engineering Research Council of Canada (NSERC) (RGPIN-2017–04464).